\documentclass[final,3p,times,authoryear]{elsarticle}

\usepackage{amssymb}

\usepackage{tikz}
\usepackage{tikz-network}
\usepackage{pgfplots}
\usepackage{subcaption}
\usepackage{multirow}
\usepackage{adjustbox}
\usepackage[font=normalsize,labelfont=bf]{caption}

\journal{Transportation Research Part C}

\begin{document}

\begin{frontmatter}



\title{Integrating On-demand Ride-sharing with Mass Transit at-Scale}

\author[label1]{Danushka Edirimanna}
\author[label1]{Hins Hu}
\author[label2]{Samitha Samaranayake}


\affiliation[label1]{
            organization={Department of Systems Engineering},
            addressline={Cornell University},
            city={Ithaca},
            country={USA}}

\affiliation[label2]{
            organization={School of Civil and Environmental Engineering},
            addressline={Cornell University},
            city={Ithaca},
            country={USA}}



\begin{abstract}
We are in the midst of a technology-driven transformation of the urban mobility landscape. However, unfortunately these new innovations are still dominated by car-centric personal mobility, which leads to concerns such as environmental sustainability, congestion, and equity. On the other hand, mass transit provides a means to move large amounts of travelers very efficiently, but is not very versatile and depends on an adequate concentration of demand. In this context, our overarching goal is to explore opportunities for new technologies such as ride-sharing to integrate with mass transit and provide a better service. More specifically, we envision a hybrid system that uses on-demand shuttles in conjunction with mass transit to move passengers efficiently, and provide an algorithmic framework for operational optimization. Our approach extends a state-of-the-art trip-vehicle assignment model to the multi-modal setting, where we develop a new integer-linear programming formulation to solve the problem efficiently. A comprehensive study covering five major cities in the United States based on real-world data is carried out to verify the advantages of such a system and the effectiveness of our algorithms. We show that our hybrid system provides significant improvements in comparison to a purely on-demand model by exploiting the efficiencies of the mass transit system. 
\end{abstract}




\end{frontmatter}


\section{Introduction}
\label{}

We are in the midst of a technology-driven transformation of the urban mobility landscape where demand-responsive ride-hailing services such as Uber and Lyft have established themselves key players. These services provide convenient and reliable access to mobility without the need for investing in a personal vehicle and the associated operating expenses such as maintenance and parking. However, recent studies have shown that ride-hailing services can increase the total number of vehicle miles traveled (VMT) in the road network compared to personal vehicle trips (~\cite{Henao2019,SCHALLER20211,doi:10.1126/sciadv.aau2670}) and also lead to an increase in the corresponding externalities, such as increased traffic congestion, greenhouse gas emissions, and transportation inequity (~\cite{doi:10.1126/sciadv.aau2670, rodier2018effects}). 

Mass transit can move passengers more efficiently (incuring lower VMT per passenger mile) by utilizing higher capacity vehicles, but is only successful in areas of dense demand~\cite{Aftabuzzaman2010EvaluatingTC}; sometimes referred to as the first-and-last-mile challenge. One strategy for mitigating this challenge is for passengers to utilize other modes such as ride-hailing and bike-sharing to connect to mass transit~\cite{shaheen2016mobility, huang2021use}.

Recently, several studies have investigated the potential benefits of integrating ride-hailing systems and carpooling services with mass transit. ~\cite{Salazar2020} showed that coordination between ride-hailing fleets and mass transit in New York City could lead to a reduction in travel times and emissions. Modeling the integration of carpooling services with mass transit, ~\cite{STIGLIC2018} showed that the overall service rate can be improved while increasing transit usage simultaneously.


\textbf{Prior work.} Many previous works on integration (~\cite{Raghunathan2023, shrivastava2006, Wang2021,  BURSTLEIN2021}) focus only on providing the first or the last mile coverage from a single transit station connected to a single transit line, often connecting a small city to a commuter rail. This simplifies the management of the ride-sharing fleet as they depart the station, drop off or pick up passengers, and return to the same station. In contrast, our research addresses a more generalized scenario with numerous transit stations (such as bus stops or subway stations) dispersed throughout a city, connecting to multiple transit lines. In this context, passengers can connect to any transit station and opt for any transit line. Under these conditions, it is not optimal to limit a ride-sharing vehicle to serve only a single transit station. Moreover, the above work only facilitates either the first mile or the last mile for a given passenger journey whereas we permit passengers to utilize both first and last-mile connectivity within the same itinerary which requires consideration of both ends when assigning passengers to ride-sharing vehicles.

In the closest study to our work, \cite{vakayil2017integrating} proposed a graph-based formulation for assigning passengers to mode-choices and managing the ride-hailing fleet to provides first-and-last mile coverage for an integrated transit system. However, their approach has the following general limitations: (1) they do not consider ride-sharing (shared rides) as an option and (2) they only consider simplistic models with respect to what transit stops each rider can transfer to and from (e.g., connections to the nearest transit stop), (3) assume average wait time for transit instead of fully integrating the transit schedule (exact waiting time depends on the passenger arrival time at the transit stop and the changing transit frequencies throughout the day), and (4) estimate passenger wait time for ride-sharing vehicles based on historical data instead of considering the current state of the ride-sharing fleet. The restrictions and assumptions, made due to the computational challenges of more general formulations, lead to limited gains from the integration and a less realistic setting for the evaluations. In particular, assumptions (3) and (4)  hinder the ability to calculate actual travel time at the time of assignment, resulting in a diminished quality of service. 


To the best of our knowledge, the existing literature does not include the study of transit-integrated ride-sharing systems through a rigorous large-scale simulation that captures microscopic movements of the transit, ride-sharing fleet, and passengers covering major metropolitan areas. 
The major cities considered in our work, such as Boston contains more than 30,000 nodes in the road network and more than 130 transit lines. In other related work, \cite{Salazar2020} studied multi-modal transportation in Berlin and New York City with a time-invariant demand modeling the behavior as a multi-commodity flow problem. As a mesoscopic study, they ignore the microscopic management of the autonomous mobility-on-demand fleet and the time-dependent attributes. 


\textbf{Contributions.} In this context, the contributions of this work are as follows: (1) Formulating the transit-integrated ride-sharing (TIRS) optimization problem for transporting passengers through a ride-sharing service that integrates with mass transit, (2) Proposing an operational optimization framework (i.e. vehicle routing and vehicle-passenger matching) using a trip-vehicle assignment model that can be solved efficiently in urban-scaled instances, (3) Validating the proposed framework via simulating the integrated ride-sharing system at-scale, and (4) Quantify the potential benefits of a transit integrated ride-sharing system in five major cities across the United States using the proposed framework.

The remainder of the paper is structured as follows. Section 2 provides a full description of the problem setting. In Section 3, the solution framework---based on an augmented request-trip-vehicle assignment model---is presented, along with techniques utilized to maintain the tractability of the problem. Numerical experiments for five major cities based on real-world data to showcase the scalability of the approach and potential benefits are given in Section 4. Finally, concluding remarks are given in Section 5.

\section{Problem Description}
\label{sec:problemDescription}

The transit-integrated ride-sharing (TIRS) problem is a real-time optimization problem that assigns passengers to multi-modal trip options and computes the corresponding routes of the ride-sharing fleet. As is common in ride-sharing operations, travel requests are aggregated over a period of time (e.g., 5-30 seconds) and an online optimization problem is solved at the corresponding frequency. All vehicles in the fleet are deployed in an on-demand manner and vehicles with passengers on board can be rerouted to serve additional passengers when capacity is available and quality of service constraints are not violated.

The objective is to fulfill as many requests as possible using a combination of ride-sharing vehicles and mass transit at the minimum cost. In this work, we assume that the cost is the total vehicle miles traveled by the ride-sharing fleet, but this can be substituted with any other metric that is linear with respect to the decision variables. We embed quality of service metrics such as waiting time and travel-time delay as hard constraints, but they can also be incorporated into the objective function.

The integrated system is restricted to serving passenger requests in the following configurations.

\begin{enumerate}
    \item \textbf{Transit option:} The mass transit provides the end-to-end connectivity. This is possible only when there is a relatively short distance between the origin/destination and the corresponding transit stops, so the passenger can walk to and from transit. When this option is available, it will always be selected over other options as it does not add costs to the system (i.e. ride-sharing vehicles are not being utilized for connectivity).
    \item \textbf{Multi-modal option:} Ride-sharing vehicles cover the first- and/or last-mile components of the journey and connect the rider with mass transit. A ride-sharing vehicle may be only needed at the origin or the destination and not both.
    \item \textbf{Ride-sharing option:} The request is fulfilled entirely by the ride-sharing vehicle. This is necessary when both of the above options are not viable due to 1) lack of mass transit connection for this request because the origin and the destination are both too far from a transit line, or 2) there are very high waiting times at the transit stop or for ride-sharing vehicles for connectivity. Note that a ride-sharing option could be selected even when a viable multi-modal option exists if the cost of the ride-sharing option is lower.  
\end{enumerate}

We make the following simplifying assumptions in the formulation that we discuss: (1) Transfers are not allowed between transit lines. We assume that travelers are unlikely to be willing to transfer transit lines in a multi-modal system. However, we will show how this assumption can be relaxed in the section 3.3. (2) The travel time of all vehicles is deterministic and known to the system. This is also consistent with the standard assumption in the ride sharing literature. We will assume that the transit system is a bus system in the rest of the article.

\begin{figure}[htbp]
    \centering
    \includegraphics[width=\linewidth]{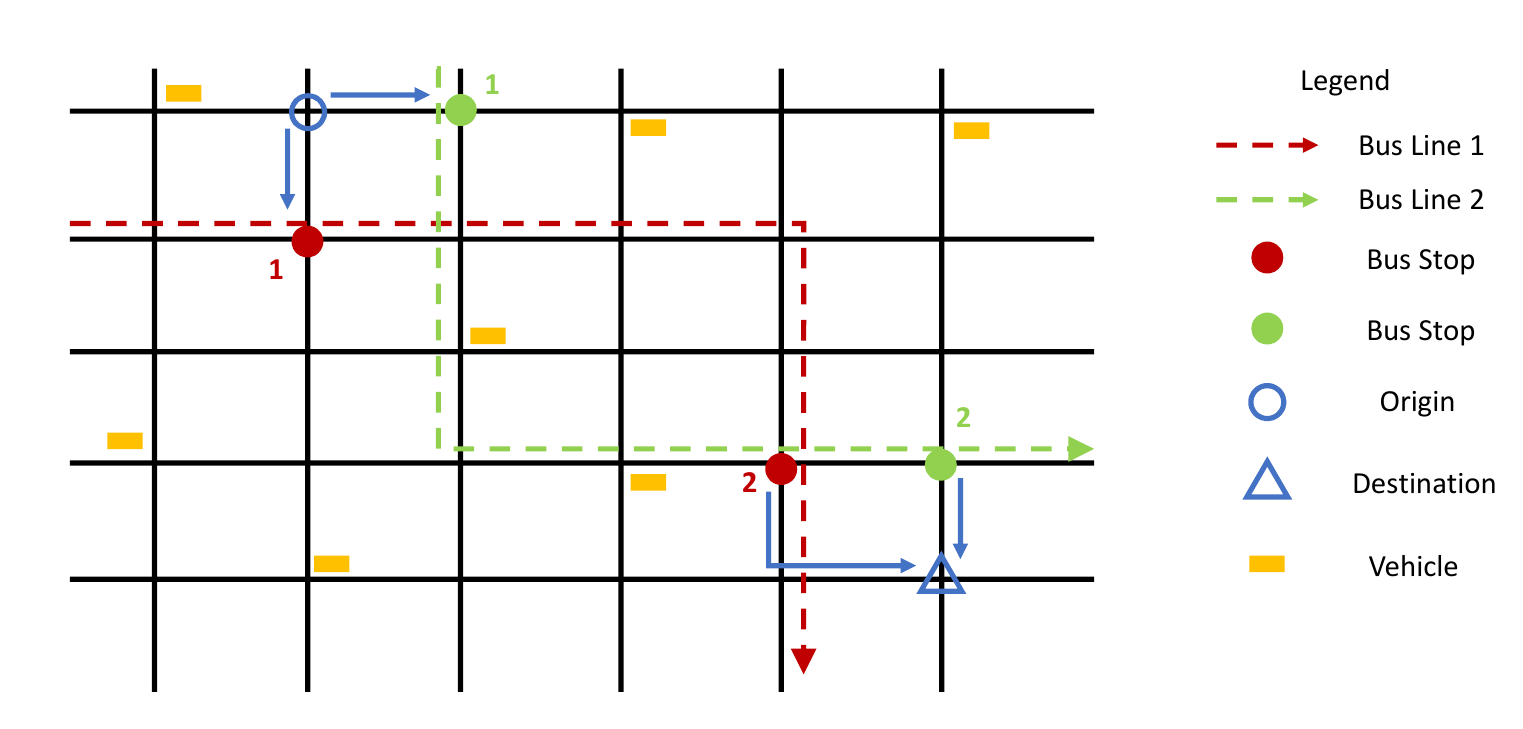}
    \caption{An illustrative example of the integrated system}
    \label{fig:illustration}
\end{figure}

Figure \ref{fig:illustration} illustrates an example, where there is one passenger request, two available bus lines, and a fleet of ride-sharing vehicles. There are many options for serving the passenger. For example, they may take a ride-sharing vehicle from the origin to stop 1 of the red line, get on the next red line bus and get off it at stop 2, and then take another ride-sharing vehicle to the destination. Alternatively, the rider may select a similar multi-modal travel option connecting to the green line.

The online problem is defined with respect to the following inputs including the system state.

\begin{enumerate}
    \item The fixed-route mass transit system specified by a collection of transit lines (routes), their schedules, and the remaining capacity of each bus.
    \item The current state of the ride-sharing vehicles specified by their current location, remaining capacity, and on-board passengers' attributes (e.g, destinations and drop-off time deadlines).
    \item The new passenger requests collected since the last passenger matching. A request specifies the earliest possible pick up time, drop off time deadline, the origin, and the final destination.
    \item The list of passengers from previous iterations who are yet to be picked up. 
\end{enumerate}

\section{The Trip-Vehicle Assignment Model}


As mentioned previously, this is an online problem that aggregates requests over a fixed \textit{batch size} and optimizes the system over each batch. Our approach is a transit-integrated extension of request-trip-vehicle (RTV) framework from Alonso-Mora et al. \cite{Alonso-Mora462}. This framework decomposes the problem into two components: (1) Trip Generation: This step consists of finding a large set of potential trip configurations and the optimal routing for each such configuration. (2) Trip-Vehicle Assignment: Once a set of candidate trip configurations are calculated we determine the optimal assignment of passengers and vehicles to trips. 

\subsection{Decomposition of the Multi-Modal Option}
As described in section 2, all requests that are satisfied by the multi-modal option could be categorized into three segments: the first mile, the transit leg, and the last mile. A transit leg specifies a combination of a departure stop, an arrival stop, a bus line, and a specific bus that operates on that line, which determines the departure and arrival times for the transit leg. There are a very large number of such combinations.

Let $L$ denote the set of bus lines in the system,  $S_l$ the set of stops located along the line $l \in L$, and $B_l$ denote the set of buses operated on line $l$. Correspondingly, $b_{li}^{da}$ represents a transit leg using bus $i$ from bus line $l$ with stop $d$ as the departure stop and stop $a$ as the arrival stop. Note that the bus identifier $i$ specifies the bus timetable (i.e., when it reaches $d$ and $a$).  $B =  \{b_{li}^{da}: \; \forall \; l \in L, \; \forall \; i \in B_l, \; \forall \; d, a \in S_l, \; d \neq a\}$ represents the set of all potential transit legs to take into account. For a single request, the size $|B|$ is $\mathcal{O} (\max\limits_{l \in L} (|L| |B_l| |S_l|^2) )$. Let $R$ be the set of passenger requests, then the total number of transit legs to consider is $\mathcal{O} (|R| \cdot \max\limits_{l \in L} (|L| |B_l| |S_l|^2))$, which is a very large number in practice. For example, in the city of Chicago, there are 67,249,306 transit leg combinations for a single request. If we accumulate 100 requests in an iteration, it would expand to 6.7 billion transit leg combinations. Furthermore, when considering ride-sharing in the next step, this number will be a factor in the size of the integer-linear program that needs to be solved to assign passenger requests to travel options and ride-sharing vehicles. Therefore, limiting the size of set B while retaining good solutions is crucial. We adopt the following pruning strategies: 
\begin{enumerate}
    \item \textbf{Exact Feasibility Checks:} A transit leg is only feasible if i) the passenger can reach the departure stop before the bus departs, ii) the passenger can reach their destination by a given deadline after getting off the bus at the arrival bus stop, and iii) there is capacity remaining in the bus from departure stop to arrival stop. We can compute a lower bound for the travel time of the first mile, which occurs when there is a ride-sharing vehicle exactly at the origin and can directly take the passenger to the bus stop. Similarly, we can compute a lower bound for the travel time of the last mile. We can use these lower bounds to filter out infeasible transit legs.
    \item \textbf{Heuristic Pruning:} Even when a transit leg is feasible, there are many such legs that are unlikely to be used in practice. For example, for a given bus line, it is most likely that passengers connect to nearby bus stops. Therefore, a stop that is far from the origin (or destination) of a request is less likely to be used if a closer stop for the same bus line exists as it will increase the length of the first (last) mile travel segment. Therefore, for each bus line, we select the bus stop that is closest to the origin of the request as the departure stop and the bus stop that is closest to the destination as the arrival bus stop. We note that this still leads to many more feasible trip configurations when only considering the nearest bus stop across all bus lines, as in~\cite{vakayil2017integrating}.
\end{enumerate}

We note that the adoption of pruning strategies is independent of our general solution framework, and that many other strategies could be employed. In this paper, our main focus is on the framework itself, and thus we introduce simple heuristics to verify computational tractability. More sophisticated pruning strategies could for example be city specific. We consider such fine-tuning to be an implementation-specific exercise.

With the aforementioned pruning strategies, the arrival stop $a \in S_l$ and departure stop $d \in S_l$ are uniquely determined for a specific bus line $l \in L$ and a request $r$. Thus, we simplify the notation $b_{li}^{a d}$ to $b_{li}$ representing a transit leg of taking bus $i$ at the closest pair of stops in line $l$ for request $r$. Accordingly, the set of candidate transit legs for one request is reduced from $B$ to a subset $\bar{B}(r) = \{b_{li}: \; \forall l \in L, \; \forall i \in B_l \}$, and thus the size of $\bar{B}(r)$ is significantly reduced to $\mathcal{O} (\sum\limits_{l \in L} |B_l| )$. For the city of Chicago, the heuristic pruning will reduce the potential number of transit legs to 17,721 for a single request, a reduction of ~3800-fold.

For each request $r \in R$ and each transit leg $b_{li} \in \bar{B}(r)$, we can generate a unique first mile $F_{r, b_{li}}$ and last mile $L_{r, b_{li}}$ segment by extending the route from the origin (or destination) to the bus departure (or arrival) stop. The set of all first mile segments is denoted by $\mathcal{F} = \{F_{r, b_{li}}: \; \forall \; r \in R, \; \forall \; b_{li} \in \bar{B}(r)\}$ and the set of all last mile segments is denoted by $\mathcal{L} = \{L_{r, b_{li}}: \; \forall \; r \in R, \; \forall \; b_{li} \in \bar{B}(r)\}$. If ride-sharing service is not necessary for the first mile and/or the last mile of a multi-modal option (e.g., passengers can walk from the origin to the departure stop), we mark this travel segment as empty. Such empty travel segments will be discussed in section 3.2 from the modeling perspective. Figure 2 visualizes the decomposition of a multi-modal option in a simple system.




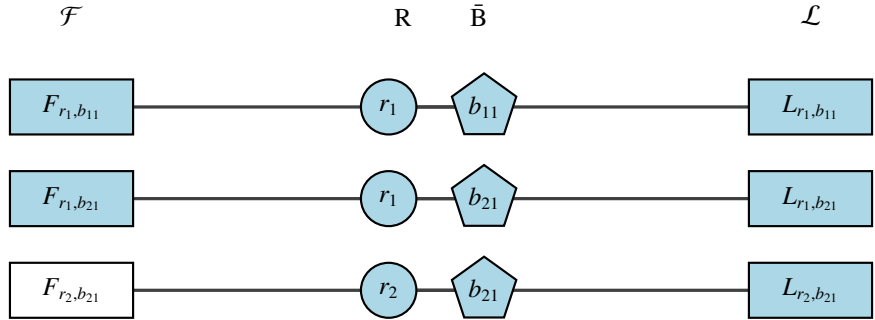
\begin{figure}[htbp]
\centering
\resizebox{0.7\linewidth}{!}{
    \begin{tikzpicture}
    \Text[x=0, y=6, fontsize=\large]{$\mathcal{F}$}
    
    \Vertex[fontsize = \large, shape=rectangle, style={minimum width=2cm}, x=0, y=4.5, size = .9, label = $F_{r_1,b_{11}}$]{f11};
    
    \Vertex[fontsize = \large, shape=rectangle, style={minimum width=2cm}, x=0, y=3, size = .9, label = $F_{r_1,b_{21}}$]{f12};
    
    \Vertex[color=white, fontsize = \large, shape=rectangle, style={minimum width=2cm}, x=0, y=1.5, size = .9, label = $F_{r_2,b_{21}}$]{f21};

    \Text[x=6, y=6, fontsize=\large]{R \qquad $\bar{\text{B}}$}
    
    \Vertex[fontsize = \large, x=5.15, y=4.5, size = .9, label = $r_1$]{r11};
    \Vertex[fontsize = \large, shape = regular polygon, size = 1.1, x=6.7, y=4.5, label = $b_{11}$]{b11};

    \Vertex[fontsize = \large, x=5.15, y=3, size = .9, label = $r_1$]{r12};
    \Vertex[fontsize = \large, shape = regular polygon, size = 1.1, x=6.7, y=3, label = $b_{21}$]{b12};

    \Vertex[fontsize = \large, x=5.15, y=1.5, size = .9, label = $r_2$]{r21};
    \Vertex[fontsize = \large, shape = regular polygon, size = 1.1, x=6.7, y=1.5, label = $b_{21}$]{b21};
    
    \Text[x=12, y=6, fontsize=\large]{$\mathcal{L}$}
    
    \Vertex[fontsize = \large, shape=rectangle, style={minimum width=2cm}, x=12, y=4.5, size = .9, label = $L_{r_1,b_{11}}$]{l11};
    
    \Vertex[fontsize = \large, shape=rectangle, style={minimum width=2cm}, x=12, y=3, size = .9, label = $L_{r_1,b_{21}}$]{l12};
    
    \Vertex[fontsize = \large, shape=rectangle, style={minimum width=2cm}, x=12, y=1.5, size = .9, label = $L_{r_2,b_{21}}$]{l21};
    
    \Edge(r11)(b11)
    \Edge(r12)(b12)
    \Edge(r21)(b21)
    
    \Edge(f11)(b11)
    \Edge(f12)(b12)
    \Edge(f21)(b21)
    
    \Edge(l11)(b11)
    \Edge(l12)(b12)
    \Edge(l21)(b21)
    
    \end{tikzpicture}
    }
\caption{The decomposition of multi-modal options in a simple system with two requests and two bus lines, where both bus lines have only one bus. Request 1 has two options, but request 2 has only one. Note that $F_{r_2,b_{21}}$ is an empty travel segment representing that passenger of the request 2 can walk to the bus stop without taking a ride-sharing vehicle}
\label{fig:rb}
\end{figure}


\subsection{Building the Transit-Integrated RTV Graph}

\begin{figure}[ht!]
\centering
\resizebox{0.8\linewidth}{0.6\textheight}{
    \begin{tikzpicture}

    \Text[x=1.5, y=13.5, fontsize=\large]{$\text{E}_1$}
    \Text[x=9, y=13.5, fontsize=\large]{$\text{E}_2$}
    
    \Text[x=-1, y=13.5, fontsize=\large]{R}
    \Vertex[fontsize = \large, x=-1, y=8, size = .9, label = $r_1$]{r1};
    \Vertex[fontsize = \large, x=-1, y=4, size = .9, label = $r_2$]{r2};

    \Text[x=5, y=13.5, fontsize=\large]{T}
    \Vertex[fontsize = \large, shape=rectangle, style={minimum width=3.0cm}, x=5, y=12, size = .9, label = $t_1 \text{ =} \{A_{r_1}\} $]{t1};

    \Vertex[fontsize = \large, shape=rectangle, style={minimum width=3.0cm}, x=5, y=10.5, size = 1, label = $t_2 \text{ =}\{F_{r_1,b_{11}} \}$]{t2};
    
    \Vertex[fontsize = \large, shape=rectangle, style={minimum width=3.0cm}, x=5, y=9, size = .9, label = $t_3 \text{ =} \{L_{r_1,b_{11}}\}$]{t3};
    
    \Vertex[fontsize = \large, shape=rectangle, style={minimum width=3.0cm}, x=5, y=7.5, size = .9, label = $t_4 \text{ =}\{F_{r_1,b_{21}} \}$]{t4};
    
    \Vertex[fontsize = \large, shape=rectangle, style={minimum width=3.0cm}, x=5, y=6, size = .9, label = $t_5 \text{ =} \{L_{r_1,b_{21}}\}$]{t5};
    
    \Vertex[fontsize = \large, shape=rectangle, style={minimum width=4.5cm}, x=5, y=4.5, size = .9, label = $t_6 \text{ =} \{L_{r_1,b_{21}} \text{,}\, L_{r_2,b_{21}}\} $]{t6};
    
    \Vertex[color=white, fontsize = \large, shape=rectangle, style={minimum width=3.0cm}, x=5, y=3, size = .9, label = $t_7 \text{ =} \{F_{r_2,b_{21}}\}$]{t7};
    
    \Vertex[fontsize = \large, shape=rectangle, style={minimum width=3.0cm}, x=5, y=1.5, size = 0.9, label = $t_8 \text{ =} \{L_{r_2,b_{21}}\}$]{t8};

    \Vertex[fontsize = \large, shape=rectangle, style={minimum width=3.0cm}, x=5, y=0, size = .9, label = $t_9 \text{ =} \{A_{r_2}\} $]{t9};

    \Text[x=13, y=13.5, fontsize=\large]{V}

    \Vertex[fontsize = \large, shape=diamond, x=13, y=10.5, size = 1.2, label = $v_1$]{v1};
    
    \Vertex[fontsize = \large, shape=diamond, x=13, y=7.5, size = 1.2, label = $v_2$]{v2};
    
    \Vertex[fontsize = \large, shape=diamond, x=13, y=4.5, size = 1.2, label = $v_3$]{v3};
    
    \Vertex[fontsize = \large, shape=diamond, x=13, y=1.5, size = 1.2, label = $v_\varnothing$]{v4};
    
    
    \Edge[opacity = .1](r1)(t1)
    \Edge[opacity = .1](r1)(t2)
    \Edge[opacity = .1](r1)(t3)
    \Edge(r1)(t4)
    \Edge[opacity = .1](r1)(t5)
    \Edge(r1)(t6)
    
    \Edge(r2)(t6)
    \Edge(r2)(t7)
    \Edge[opacity = .1](r2)(t8)
    \Edge[opacity = .1](r2)(t9)

    \Edge[opacity = .1](t1)(v1)
    \Edge[opacity = .1](t2)(v1)
    \Edge[opacity = .1](t3)(v1)
    
    \Edge[label = $c_{t_4\text{,}v_1}$, fontsize = \large](t4)(v1)
    \Edge[label = $c_{t_6\text{,}v_2}$, fontsize = \large](t6)(v2)
    \Edge[label = $0$, fontsize = \large](t7)(v4)

    \Edge[opacity = .1](t4)(v2)
    \Edge[opacity = .1](t5)(v2)
    \Edge[opacity = .1](t6)(v2)
    \Edge[opacity = .1](t7)(v2)

    \Edge[opacity = .1](t5)(v3)
    \Edge[opacity = .1](t6)(v3)
    \Edge[opacity = .1](t8)(v3)
    \Edge[opacity = .1](t9)(v3)
    
    \end{tikzpicture}
    }
\caption{The Transit-Integrated RTV graph of a small system with two requests, nine potential travel options, and three vehicles. Request $1$ rides on vehicle $1$ to catch the bus $b_{21}$ and rides on vehicle $2$ for the last mile. Vehicle $3$ covers the last mile of the request $2$. Note that the first mile of request 2 is matched to the dummy vehicle because it is an empty (walking) travel segment. Vehicle $2$ serves as both the last-mile vehicle of the request $2$ and the last-mile vehicle of the request $1$. The other vehicles only support one travel segment each.}
\label{fig:rtv}
\end{figure}
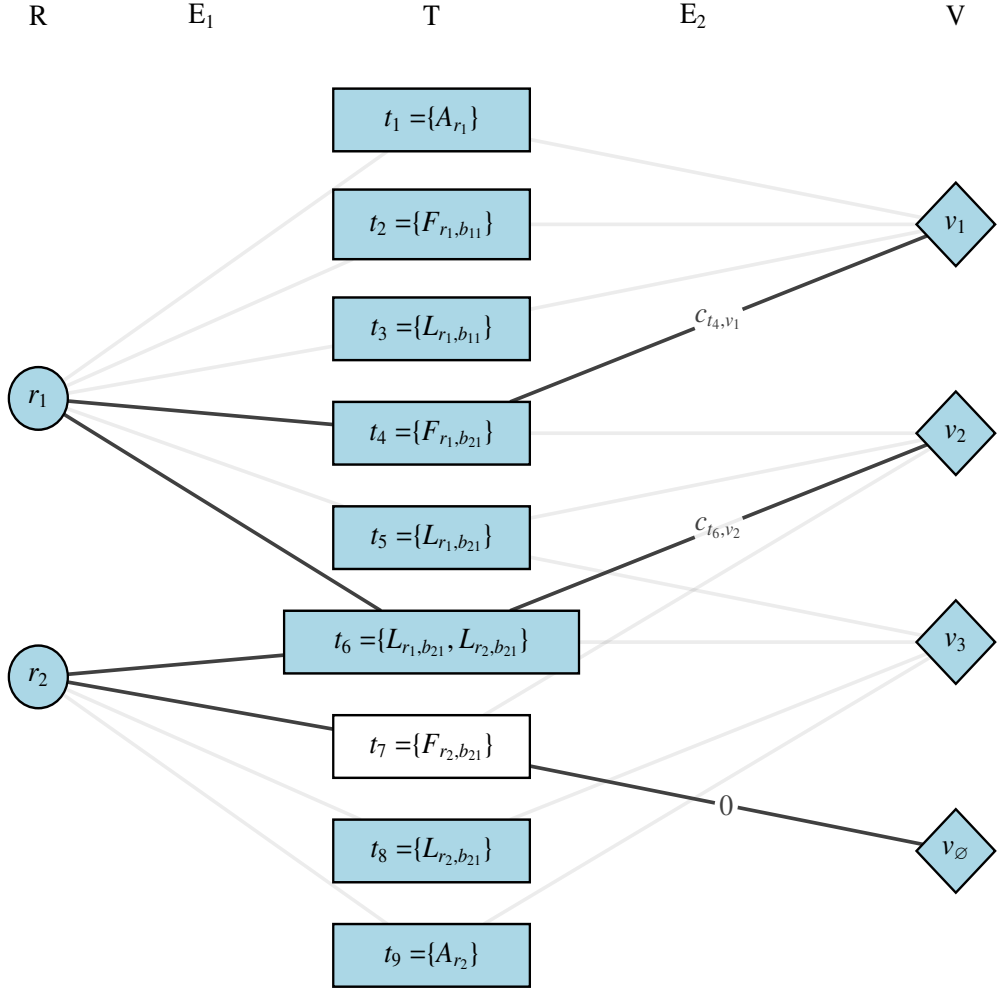

As proposed by Alonso-Mora et al. \cite{Alonso-Mora462}, the RTV graph captures the relationship among a list of passenger requests $R$, a list of trips $T$, and a list of ride-sharing vehicles $V$. In the RTV framework, a trip $t \in T$ is defined as a set of requests that can be potentially served by a single vehicle, since there are no transit integrated trips in their setting. A pairwise compatibility graph, also known as a shareability graph (originally defined in~\cite{Santi13290}) is first created. An edge connecting two travel segments indicates that a vehicle can serve them simultaneously without violating any of the pre-defined QoS (Quality of Service) constraints (e.g., waiting time and detour), and an edge connecting a request and a vehicle indicates there is no constraint violated by matching them together. Each feasible trip must correspond to a clique (i.e., a complete sub-graph) in the shareability graph. Note that this is a necessary but not sufficient condition for trips of size 3 or larger. In that case, a generalized TSP or capacitated pickup and delivery problem with time windows needs to be solved to determine feasibility and obtain the optimal route. 

In this work, we extend the RTV framework to a new transit-integrated RTV (TI-RTV) model. A trip (T in RTV) in this setting corresponds to a collection of travel segments taken by some requests as opposed to the entirety of their travel. A travel segment can be a first mile, a last mile, or an entire route between the origin and the destination that is served by a ride-sharing vehicle. 


The construction of TI-RTV graph is as follows. First, we initialize the list of travel segments as $T = \mathcal{F} \bigcup \mathcal{L}$ by adding all first-mile and last-mile travel segments. In addition, we also include the travel segment $A_r~ \forall r \in R$ from the ride-sharing only option, which expands the list to $T = T \; \bigcup \; \{A_r: \; \forall \; r \in R\}$. Then, we construct the travel segment-travel segment sharability graph similar to the request-request sharability graph in the RTV framework by adding feasible cliques. However, checking whether a clique is feasible is stricter in the TI-RTV setting due to additional physical constraints. Particularly, the following travel segments are mutually exclusive and hence cannot share the same vehicle: (1) Multiple first mile segments arising from the same request, (2) Multiple last mile segments arising from the same request, and (3) The first mile and the last mile arising from the same request but connecting different transit legs. While in theory there could be a very large number of feasible shared travel segments, due to relatively strict QoS constraints (e.g., maximum waiting time and detour) in practice, the total number of feasible shared segments is manageable in practice. 



Once we have the finalized list $T$, the next step is to match $R$ to $T$ and match $T$ to $V$. We define a bipartite graph $G_1 = (R \; \cup \; T, \; E_1)$ where $E_1$ is the set of edges corresponding to travel segments in a $t \in T$ that arise from a request $r \in R$.  Define another bipartite graph $G_2 = (T \; \cup \; V, \; E_2)$, where $E_2$ corresponds to dispatching a vehicle $v \in V$ to a trip $t \in T$ being feasible. Each associated weight in $E_2$ is the increased cost of the vehicle serving the trip. 

Finally, to model the empty travel segments corresponding to the first miles or the last miles, a dummy vehicle $v_{\varnothing}$ is added to the list $V$ connecting all of these empty travel segments and zero weights are assigned to the edges. It is justified because (1) matching them induces zero cost to the objective and (2) it avoids assigning regular vehicles to unnecessary first miles and last miles. Figure 3 visualizes a TI-RTV graph of a small system.

In the context of ride-sharing, computing the cost or feasibility for a vehicle $v \in V$ to serve all requests in a trip $t \in T$ involves solving a pickup and delivery problem (PDP). However, we do not need to run the PDP for each $v-t$ combination. If $t$ contains more than one travel segment, we know that each subset of $t$ is also an entry in the list $T$. We need to compute the feasibility or the cost of a $v-t$ combination if and only if a vehicle $v$ can serve each such subset. In other words, a $v-t$ combination is feasible if and only if there is an edge connecting $v$ and all subsets of $t$ in the RTV graph, which makes it more efficient to perform computation in an increasing order of the size of $t$. Moreover, the computation of a $v-t$ edge only depends on the particular vehicle $v$, which allows the PDPs to be solved in parallel.


We note that in our formulation, a ride-sharing vehicle may only partially serve a request (i.e., either the first-mile segment or the last-mile segment of a multi-modal option). Therefore, matching a request to a single node in $T$ may not be sufficient.




\subsection{Vehicle Assignment via Integer-Linear Programming}

In this section, we formulate an integer-linear program (ILP) to determine the optimal trip-vehicle assignment based on the TI-RTV graph. Let $x_{tv}$ denote a binary variable indicating whether trip $t$ is assigned to vehicle $v$. Let $y_{rt}$ denote a binary variable indicating whether request $r$ takes the trip $t$. If a request $r$ is not served, another binary variable $z_{r}$ is set to 1. Otherwise, it is set to 0. Let $F(r)$ denote the subset of $T$ corresponding to all trips that contain a first-mile travel segment of request $r$. Similarly, denote by $L(r)$ the subset of $T$ corresponding to all trips that contain a last-mile travel segment of request $r$. Let $A(r)$ denote the set of all trips that serve the request $r$ with the ride-sharing option. Recall that $F_{r, b_{li}}$ and $L_{r, b_{li}}$ represent the first-mile and the last-mile travel segments of request $r$ if the rider takes bus trip $b_{li}$ for $l \in L$ and $i \in B_l$, and that $\bar{B}(r)$ is the set of candidate transit legs for request $r$. To make the notations more concise, we simplify $F_{r, b_{li}}$ and $L_{r, b_{li}}$ to $F_{r, b}$ and $L_{r, b}$ for $\forall \; b \in \bar{B}(r)$. Given the above definitions, we formulate the ILP as follows.

\begin{align}
    \min_{x,y,z} \quad &\sum_{v \in V, \; t \in T} c_{tv} \; x_{tv} + \sum_{r \in R} p_r \; z_{r}  \\
    s.t. \quad & \sum_{t \in T} x_{tv} \leq 1 \qquad \forall \; v \in V\backslash\{v_{\varnothing}\} \\
    & \hspace{-0.15cm} \sum_{t \in F(r)} y_{rt} + \sum_{t \in A(r)} y_{rt} + z_{r} = 1 \qquad \forall \; r \in R \\
    & \hspace{-0.6cm} \sum_{t \in F(r): \, F_{r, b} \in t} y_{rt} - \sum_{t \in L(r): \, L_{r, b} \in t} y_{rt} = 0 \qquad \forall \; r \in R, \; \forall \; b \in \bar{B}(r)\\
    & \sum_{v \in V} x_{tv} = y_{rt} \qquad \forall \; r \in R, \; \forall \; t \in T \\
    & x_{tv}, \; y_{rt}, \; z_{r} \in \{0, 1\} \\
\end{align}

Equation (1) defines the objective function, where the first term is the cost of all ride-sharing vehicles serving all trips (with pre-computed costs) and the second term is a penalty term for all requests that are not served. Constraint (2) regulates that a vehicle can only be assigned one trip at most except for the dummy vehicle. Constraint (3) enforces that a single travel option is selected for each request (either a multi-modal trip or a ride-sharing only trip) if the request is served. Constraint (4) guarantees consistency between the first mile and the last mile segments. In particular, if a trip is multi-modal, both the first mile and the last mile must be assigned to ride-sharing vehicles even though they may be empty travel segments as marked in the TI-RTV graph. Constraint (5) enforces that every trip $t \in T$ that is served must be assigned to a vehicle. The total number of constraints is $\mathcal{O}(|V| + |R| + |T|)$ and the number of decision variables is $\mathcal{O}(|V||T|+|R||T|)$, where $|T| = \mathcal{O}(|R||\bar{B}|)$. 


We only consider transit legs if there is capacity remaining in the relevant bus from the departure stop to the arrival stop (Exact feasibility check). However, in the constraints, we do not capture the potential bus capacity conflicts among requests. In order for bus capacity to be violated with such conflicts, 1) the transit legs of conflicting requests should be overlapped (traveling in the same bus on the same line and in between overlapping bus stops) and 2) the bus should be close to its full capacity (the number of conflicting transit legs should be greater than the remaining capacity). However, since the number of requests in a batch (aggregated within 30 seconds) is very small, such conflicts are unlikely, and potential violations are very small.


\textbf{Allowing multi-modal trips with bus-to-bus transfers }
Note that we made two simplifying assumptions for the TIRS problem in Section \ref{sec:problemDescription}, which improved the computational complexity of constructing the trip-vehicle assignment model. We can relax the no multi-modal trips with bus-to-bus transfers assumption at some additional computational cost, while the solution framework intact. This can be done by extending the set $B$ to contain bus trips involving more than one bus. A multi-bus transit leg is similar to a single-bus transit leg except that we need to determine the transfer stops and consider the transfer time in the feasibility check. Then, the subsequent workflow is identical to what we discussed in previous sections. In practice, it is unlikely that more than one bus-to-bus transfer is viable in a multi-modal system due to the potential first and last-mile connections.


\section{Numerical Experiments}
In this section, we simulate an integrated system using real-world data to assess the performance of our trip-assignment model. 

\subsection{Settings}

We simulate commuter trips for five major cities across unites states; Atlanta, Boston, Chicago, Houston, and Los Angeles. To generate the commuter trip demand, the LEHD Origin-Destination Employment Statistics (LODES) data \cite{lodes}, released by the United States Census Bureau, serves as the foundational dataset, providing home-work locations of employees at the Census Block level. A Census Block is generally defined as a city block bordered by surrounding streets. The process of creating Origin-Destination pairs of demand entails the random selection of both home and workplace locations within the designated Census block. In order to reflect realistic commuting patterns, morning trips are uniformly distributed between 6 am and 8 am, while evening trips are distributed from 4 pm to 6 pm. We filter out trips that are shorter than 3km as they are less likely to be suited for multi-modal travel.

The road network and travel time matrix are obtained from OpenStreetMap \cite{openstreetmap}. Our implementation of the trip simulator requires loading a pre-computed travel time matrix of the city into the memory. Our server can accommodate up to matrices for ~32,000 nodes. Consequently, the coverage of each city is constrained to a circular region around centering the downtown area, with a radius of 12km for Boston, Houston, and Los Angeles, and 15km and 16km for Atlanta and Chicago, respectively. To expedite the experiments, a 10\% data sample is utilized while maintaining data consistency across multiple settings.

We rely on General Transit Feed Specification (GTFS) data published by respective transit agencies for the transit schedule. The simulator is written in Python (3.9.7) and the ILP is solved using the commercial solver Gurobi (9.5.1). All the experiments are performed on a Linux Server equipped with Intel(R) Xeon(R) Gold 6244 CPU @ 3.60GHz processor and 200 GB of memory.

Each request  $r \in R$ is subject to a maximum travel duration constraint $T_{max} (r)$, defined as the summation of a factor of the direct shortest path travel time from the origin to the destination of the request ($SPTT(r)$) and a constant additional time ($\beta$), as specified in equation (7). The parameter $\alpha$ represents a multiplier that models the extent to which individuals are willing to tolerate extra time spent on the road, accounting for detours in ride-sharing and multi-modal travel scenarios. Additionally, a constant time value, $\beta$, is introduced to account for waiting periods associated with transit and ride-sharing vehicle arrivals. In the experiments, we set $\alpha$ to $0.2$ and $\beta$ to $20$ minutes.

\begin{align}
T_{max} (r) &= (1+\alpha) * SPTT(r) + \beta
\end{align}

To reduce computation times, the matching algorithm in the numerical experiments only assigns at most one request to each vehicle in each batch (aggregated within 30 seconds or up to 100 requests, whichever is completed earliest). Note that this does not prevent ride-sharing as multiple riders can be assigned to the same vehicle in successive batches. Extensive independent experiments by Simonetto et al. \cite{Simonetto2019} have shown that this assumption does not degrade performance significantly. This simplification can be eliminated with additional computational resources.

To evaluate the proposed integrated system, we carry out the simulations under three different settings; 1) serving requests with both ride-sharing and multi-modal options (fully integrated system), 2) serving requests only with the ride-sharing option, and 3) serving the requests with only the multi-modal option. The comparison of these scenarios will be shown in the next section. In the experiments, we discard the trips that can be served only using the transit as they can be served independently from the ride-sharing fleet.

\subsection{Results}

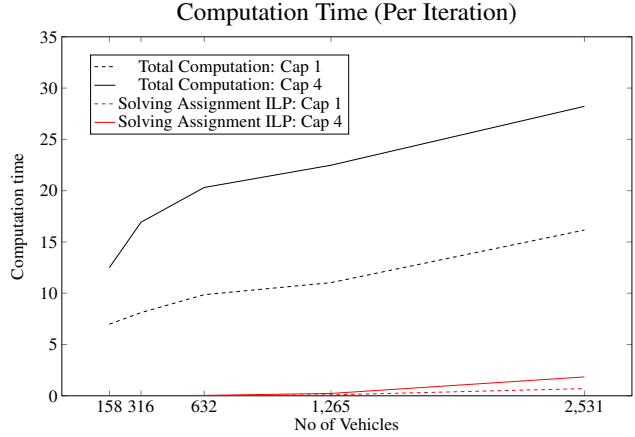
\begin{figure*}[t!]
    \centering
    \begin{adjustbox}{width=.5\textwidth}
\begin{tikzpicture}
\begin{axis}
[
ymin = 0, ymax=35,
xtick = data,
title={Computation Time (Per Iteration)},
width=18cm,
height=12cm,
legend style={
            font=\Large,
            at={(0.05,0.95)},
            anchor= north west,
        },
tick label style={font=\Large}, 
label style={font=\Large},
title style={font=\huge},
ylabel=Computation time,
xlabel=No of Vehicles,
]
\addplot[color=black, style=dashed] table [x, y=Total C1, col sep=comma] {graphs/csv/exe_time_per_iteration.csv};
\addplot[color=black] table [x, y=Total C4, col sep=comma] {graphs/csv/exe_time_per_iteration.csv};
\addplot[color=red, style=dashed] table [x, y=Optimization C1, col sep=comma] {graphs/csv/exe_time_per_iteration.csv};
\addplot[color=red] table [x, y=Optimization C4, col sep=comma] {graphs/csv/exe_time_per_iteration.csv};
\legend{Total Computation: Cap 1, Total Computation: Cap 4, Solving Assignment ILP: Cap 1,Solving Assignment ILP: Cap 4}

\end{axis}
\end{tikzpicture}
\end{adjustbox}
    \caption{Average computation time of the proposed framework (per iteration with 100 requests) for the city of Chicago: Comparison with varying vehicle capacities (1 and 4) and fleet sizes (2.5, 5, 10, 20, 40 per 1000 requests)}
      \label{fig:exe_time}
\end{figure*}

\begin{figure*}[p!]
    \input{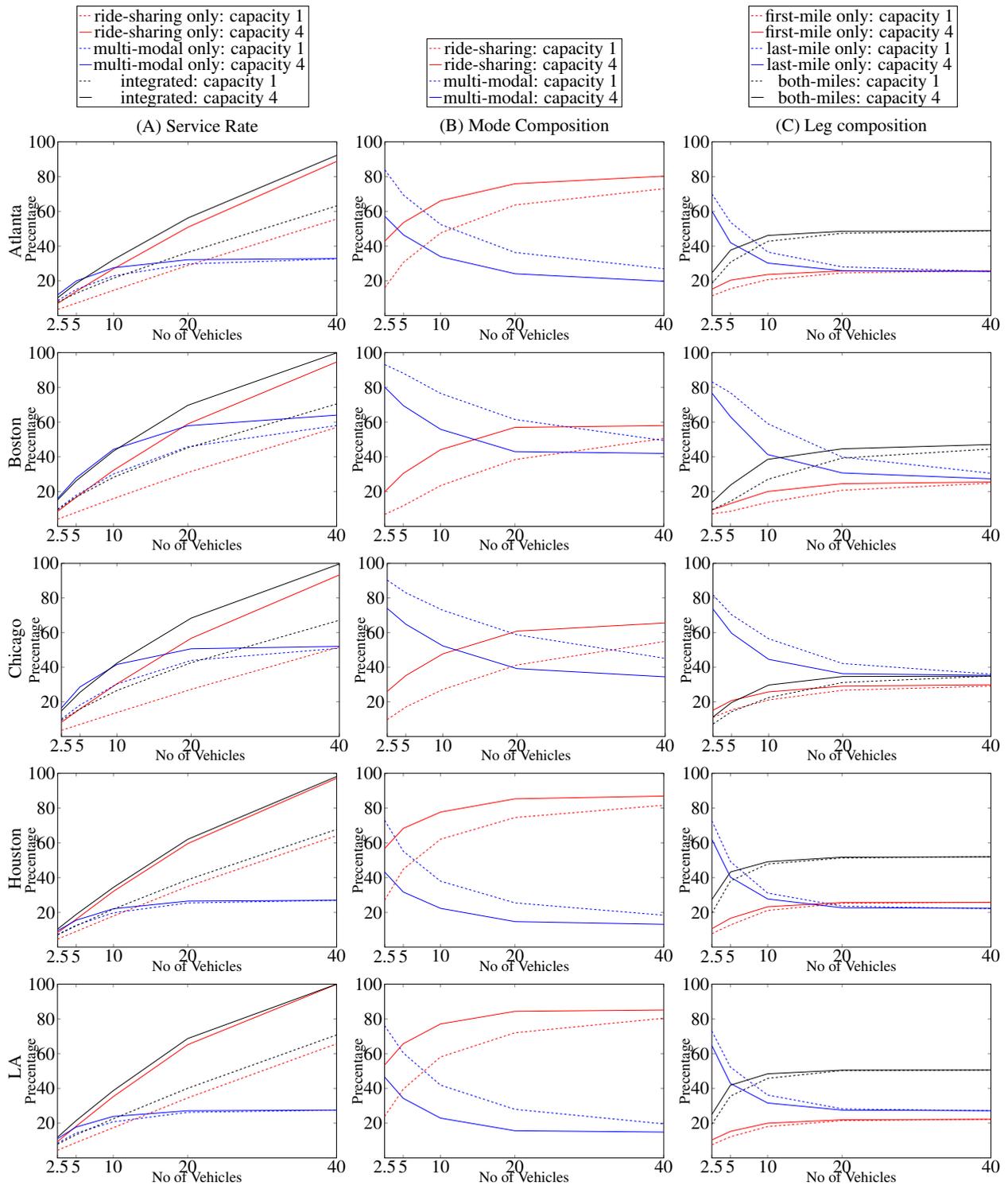}
    \caption{Comparison of different metrics for varying vehicle capacities (1 and 4) and fleet sizes (2.5, 5, 10, 20, 40 per 1000 requests). The rows represent results for different cities while the columns contain results for different metrics. (A) Comparison of service rate, (B) Comparison of the composition of service options, and (C) Comparison of the distribution of types of multi-modal trips.}
      \label{fig:results}
\end{figure*}

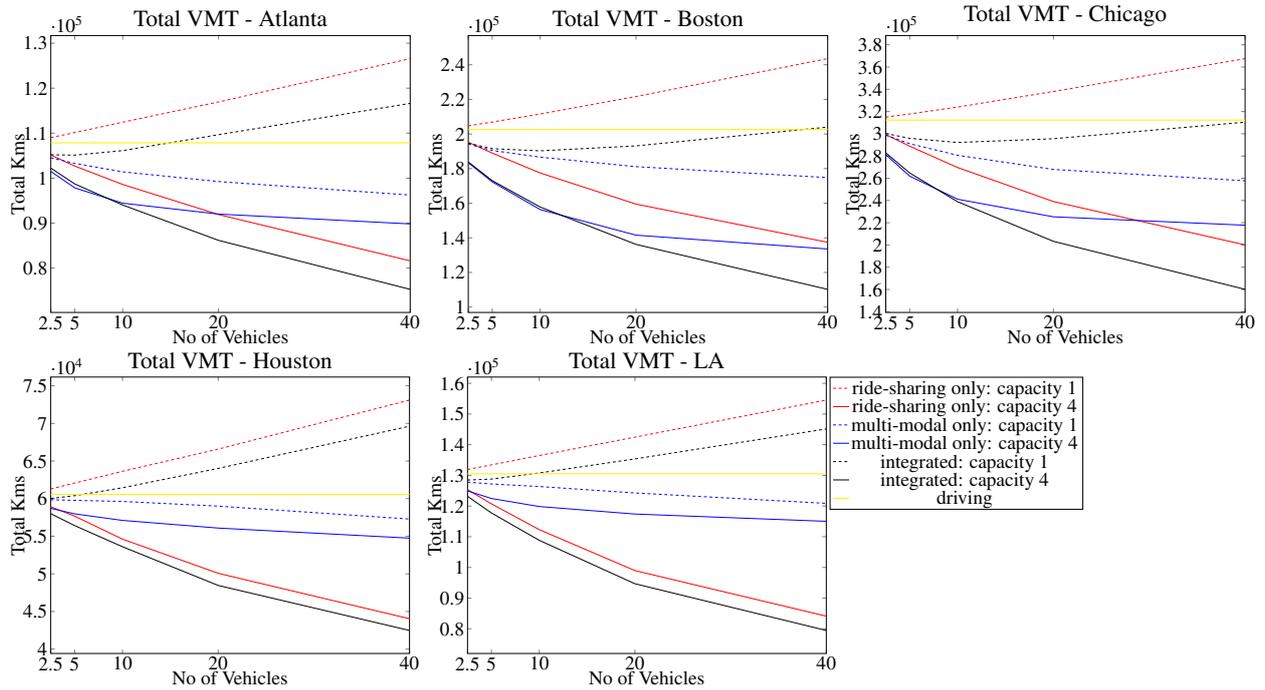
\begin{figure*}
    \begin{adjustbox}{width=\textwidth}
\begin{tikzpicture}
\begin{axis}
[
xmin = 2.5, xmax=40,
xtick = data,
width=15cm,
height=12cm,
title={Total VMT - Atlanta},
legend style={
            font=\huge,
            at={(1,1.2)},
            anchor=south east,
        },
tick label style={font=\huge}, 
label style={font=\huge},
title style={font=\Huge},
ylabel=Total Kms,
xlabel=No of Vehicles,
]
\addplot[color=red, style=dashed] table [x, y=Atlanta OD Cap 1, col sep=comma] {graphs/csv/vmt.csv};
\addplot[color=red] table [x, y=Atlanta OD Cap 4, col sep=comma] {graphs/csv/vmt.csv};
\addplot[color=blue, style=dashed] table [x, y=Atlanta Multi-Modal Cap 1, col sep=comma] {graphs/csv/vmt.csv};
\addplot[color=blue] table [x, y=Atlanta Multi-Modal Cap 4, col sep=comma] {graphs/csv/vmt.csv};
\addplot[color=black, style=dashed] table [x, y=Atlanta Integrated Cap 1, col sep=comma] {graphs/csv/vmt.csv};
\addplot[color=black] table [x, y=Atlanta Integrated Cap 4, col sep=comma] {graphs/csv/vmt.csv};
\addplot[color=yellow] table [x, y=Atlanta Driving, col sep=comma] {graphs/csv/vmt.csv};

\end{axis}
\end{tikzpicture}
\begin{tikzpicture}
\begin{axis}
[
xmin = 2.5, xmax=40,
xtick = data,
width=15cm,
height=12cm,
title={Total VMT - Boston},
legend image code/.code={%
                \draw[#1, draw=none] (0cm,-0.1cm) rectangle (0.6cm,0.1cm);
            },
legend style={
            font=\small,
            at={(0.1,.85)},
            anchor=west,
        },
tick label style={font=\huge}, 
label style={font=\huge},
title style={font=\Huge},
ylabel=Total Kms,
xlabel=No of Vehicles,
]
\addplot[color=red, style=dashed] table [x, y=Boston OD Cap 1, col sep=comma] {graphs/csv/vmt.csv};
\addplot[color=red] table [x, y=Boston OD Cap 4, col sep=comma] {graphs/csv/vmt.csv};
\addplot[color=blue, style=dashed] table [x, y=Boston Multi-Modal Cap 1, col sep=comma] {graphs/csv/vmt.csv};
\addplot[color=blue] table [x, y=Boston Multi-Modal Cap 4, col sep=comma] {graphs/csv/vmt.csv};
\addplot[color=black, style=dashed] table [x, y=Boston Integrated Cap 1, col sep=comma] {graphs/csv/vmt.csv};
\addplot[color=black] table [x, y=Boston Integrated Cap 4, col sep=comma] {graphs/csv/vmt.csv};
\addplot[color=yellow] table [x, y=Boston Driving, col sep=comma] {graphs/csv/vmt.csv};

\end{axis}
\end{tikzpicture}
\begin{tikzpicture}
\begin{axis}
[
xmin = 2.5, xmax=40,
xtick = data,
width=15cm,
height=12cm,
title={Total VMT - Chicago},
legend image code/.code={%
                \draw[#1, draw=none] (0cm,-0.1cm) rectangle (0.6cm,0.1cm);
            },
legend style={
            font=\small,
            at={(0.1,.85)},
            anchor=west,
        },
tick label style={font=\huge}, 
label style={font=\huge},
title style={font=\Huge},
ylabel=Total Kms,
xlabel=No of Vehicles,
]
\addplot[color=red, style=dashed] table [x, y=Chicago OD Cap 1, col sep=comma] {graphs/csv/vmt.csv};
\addplot[color=red] table [x, y=Chicago OD Cap 4, col sep=comma] {graphs/csv/vmt.csv};
\addplot[color=blue, style=dashed] table [x, y=Chicago Multi-Modal Cap 1, col sep=comma] {graphs/csv/vmt.csv};
\addplot[color=blue] table [x, y=Chicago Multi-Modal Cap 4, col sep=comma] {graphs/csv/vmt.csv};
\addplot[color=black, style=dashed] table [x, y=Chicago Integrated Cap 1, col sep=comma] {graphs/csv/vmt.csv};
\addplot[color=black] table [x, y=Chicago Integrated Cap 4, col sep=comma] {graphs/csv/vmt.csv};
\addplot[color=yellow] table [x, y=Chicago Driving, col sep=comma] {graphs/csv/vmt.csv};

\end{axis}
\end{tikzpicture}
\end{adjustbox}

\begin{adjustbox}{width=0.86\textwidth}

\begin{tikzpicture}
\begin{axis}
[
xmin = 2.5, xmax=40,
xtick = data,
width=15cm,
height=12cm,
title={Total VMT - Houston},
legend image code/.code={%
                \draw[#1, draw=none] (0cm,-0.1cm) rectangle (0.6cm,0.1cm);
            },
legend style={
            font=\small,
            at={(0.1,.85)},
            anchor=west,
        },
tick label style={font=\huge}, 
label style={font=\huge},
title style={font=\Huge},
ylabel=Total Kms,
xlabel=No of Vehicles,
]
\addplot[color=red, style=dashed] table [x, y=Houston OD Cap 1, col sep=comma] {graphs/csv/vmt.csv};
\addplot[color=red] table [x, y=Houston OD Cap 4, col sep=comma] {graphs/csv/vmt.csv};
\addplot[color=blue, style=dashed] table [x, y=Houston Multi-Modal Cap 1, col sep=comma] {graphs/csv/vmt.csv};
\addplot[color=blue] table [x, y=Houston Multi-Modal Cap 4, col sep=comma] {graphs/csv/vmt.csv};
\addplot[color=black, style=dashed] table [x, y=Houston Integrated Cap 1, col sep=comma] {graphs/csv/vmt.csv};
\addplot[color=black] table [x, y=Houston Integrated Cap 4, col sep=comma] {graphs/csv/vmt.csv};
\addplot[color=yellow] table [x, y=Houston Driving, col sep=comma] {graphs/csv/vmt.csv};

\end{axis}
\end{tikzpicture}
\begin{tikzpicture}
\begin{axis}
[
xmin = 2.5, xmax=40,
xtick = data,
width=15cm,
height=12cm,
title={Total VMT - LA},
legend style={
            font=\huge,
            at={(1.01,1)},
            anchor=north west,
        },
tick label style={font=\huge}, 
label style={font=\huge},
title style={font=\Huge},
ylabel=Total Kms,
xlabel=No of Vehicles,
]
\addplot[color=red, style=dashed] table [x, y=LA OD Cap 1, col sep=comma] {graphs/csv/vmt.csv};
\addplot[color=red] table [x, y=LA OD Cap 4, col sep=comma] {graphs/csv/vmt.csv};
\addplot[color=blue, style=dashed] table [x, y=LA Multi-Modal Cap 1, col sep=comma] {graphs/csv/vmt.csv};
\addplot[color=blue] table [x, y=LA Multi-Modal Cap 4, col sep=comma] {graphs/csv/vmt.csv};
\addplot[color=black, style=dashed] table [x, y=LA Integrated Cap 1, col sep=comma] {graphs/csv/vmt.csv};
\addplot[color=black] table [x, y=LA Integrated Cap 4, col sep=comma] {graphs/csv/vmt.csv};
\addplot[color=yellow] table [x, y=LA Driving, col sep=comma] {graphs/csv/vmt.csv};

\legend{ride-sharing only: capacity 1,ride-sharing only: capacity 4,multi-modal only: capacity 1,multi-modal only: capacity 4, integrated: capacity 1, integrated: capacity 4,driving}

\end{axis}
\end{tikzpicture}

\end{adjustbox}
    \caption{Total Vehicle Miles Traveled (VMT) for varying vehicle capacities (1 and 4) and fleet sizes (2.5, 5, 10, 20, 40 per 1000 requests).}
      \label{fig:vmt}
\end{figure*}

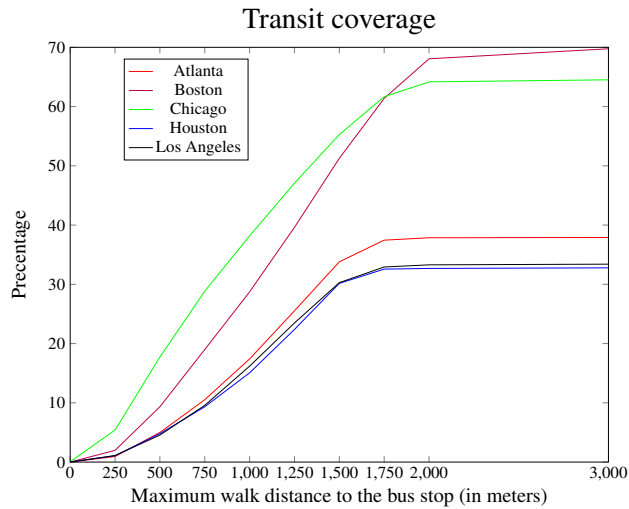
\begin{figure*}[t!]
    \centering
    \begin{adjustbox}{width=0.5\textwidth}
\begin{tikzpicture}
\begin{axis}
[
ymin = 0, ymax=70,
xmin = 0, xmax=3000,
xtick = data,
title={Transit coverage},
width=15cm,
height=12cm,
legend style={
            font=\large,
            at={(0.1,.85)},
            anchor=west,
        },
tick label style={font=\large}, 
label style={font=\Large},
title style={font=\huge},
ylabel=Precentage,
xlabel=Maximum walk distance to the bus stop (in meters),
]
\addplot[color=red] table [x, y=Atlanta, col sep=comma] {graphs/csv/transit_only_vs_walk_distance.csv};
\addplot[color=purple] table [x, y=Boston, col sep=comma] {graphs/csv/transit_only_vs_walk_distance.csv};
\addplot[color=green] table [x, y=Chicago, col sep=comma] {graphs/csv/transit_only_vs_walk_distance.csv};
\addplot[color=blue] table [x, y=Houston, col sep=comma] {graphs/csv/transit_only_vs_walk_distance.csv};
\addplot[color=black] table [x, y=LA, col sep=comma] {graphs/csv/transit_only_vs_walk_distance.csv};

\legend{Atlanta, Boston, Chicago, Houston, Los Angeles}
\end{axis}
\end{tikzpicture}
\end{adjustbox}
    \caption{Transit Reach vs Walking Distance: Percentage of commuters that can commute only using the transit if they are willing to walk given distance to the stop.}
      \label{fig:transit_only}
\end{figure*}

We simulate the ride-sharing fleet with varying capacities (1 and 4) and fleet sizes. We consider 5 different fleet sizes for each city proportionate to the number of passenger requests as specified in Table \ref{table:1}.

\begin{table*}[]
\centering
\begin{tabular}{|l|l|l|llllll|}
\hline
\multirow{2}{*}{City} & \multirow{2}{*}{No of requests} & \multirow{2}{*}{Area Radius} & \multicolumn{6}{l|}{Number of vehicles}                                                                                                                                 \\ \cline{4-9} 
                      &                                 &                              & \multicolumn{1}{l|}{2.5} & \multicolumn{1}{l|}{5}   & \multicolumn{1}{l|}{10}  & \multicolumn{1}{l|}{20}   & \multicolumn{1}{l|}{40}   & Per 1000 requests              \\ \hline
Atlanta               & 22138                           & 15km                         & \multicolumn{1}{l|}{55}  & \multicolumn{1}{l|}{110} & \multicolumn{1}{l|}{221} & \multicolumn{1}{l|}{442}  & \multicolumn{1}{l|}{885}  & \multirow{5}{*}{Actual Number} \\ \cline{1-8}
Boston                & 44766                           & 12km                         & \multicolumn{1}{l|}{112} & \multicolumn{1}{l|}{224} & \multicolumn{1}{l|}{448} & \multicolumn{1}{l|}{895}  & \multicolumn{1}{l|}{1790} &                                \\ \cline{1-8}
Chicago               & 63283                           & 16km                         & \multicolumn{1}{l|}{158} & \multicolumn{1}{l|}{316} & \multicolumn{1}{l|}{632} & \multicolumn{1}{l|}{1265} & \multicolumn{1}{l|}{2531} &                                \\ \cline{1-8}
Houston               & 14687                           & 12km                         & \multicolumn{1}{l|}{36}  & \multicolumn{1}{l|}{73}  & \multicolumn{1}{l|}{147} & \multicolumn{1}{l|}{294}  & \multicolumn{1}{l|}{588}  &                                \\ \cline{1-8}
LA                    & 30728                           & 12km                         & \multicolumn{1}{l|}{77}  & \multicolumn{1}{l|}{154} & \multicolumn{1}{l|}{307} & \multicolumn{1}{l|}{615}  & \multicolumn{1}{l|}{1230} &                                \\ \hline
\end{tabular}
\caption{Settings for cities.}
\label{table:1}
\end{table*}

The simulation includes various configurations of the ride-sharing fleet, with distinct capacities (1 and 4) and differing fleet sizes (ranging from 2.5 to 40 vehicles per 1000 requests), as outlined in Table 1. In the transit lines, the capacity of busses is set at 50, while other lines such as subways and commuter rails are set to a capacity of 1000.

Figure \ref{fig:exe_time} presents the average computational time required to solve a single iteration of the proposed integrated transit system with 100 passenger requests in Chicago. With a total of 63,283 passengers and up to 2,531 ride-sharing vehicles in the city, Chicago is chosen for representation due to its larger scale. This figure demonstrates the real-time feasibility of the proposed framework. The 63,283 commuter requests are distributed over a 4-hour period, equating to 263 requests per minute. Therefore, the framework, on average, needs to execute 2.63 times per minute. The integrated system consistently completes these iterations in under one minute for all configurations, except when utilizing 2531 ride-sharing vehicles, where it extends slightly by an extra 14 seconds. However, we observe that the bulk of the computation time (more than 95\%) is spent on building the transit-integrated RTV graph which can be built in parallel. Therefore, we can further reduce the computation time with more computing resources and achieve real-time response. Additionally, Figure \ref{fig:exe_time} illustrates that the system exhibits linear scalability with respect to the number of ride-sharing vehicles employed in the simulation.

Column A of Figure \ref{fig:results} depicts the resultant service rates across each configuration. A consistent trend emerges across all cities. Notably, the integrated setting consistently achieves higher service rates than the ride-sharing-only setting. However, the extent of service rate enhancement varies by city. Specifically, with a capacity of 4, Atlanta, Houston, and LA display a marginal absolute improvement in service rate at 5.34\%, 2.78\%, and 3.52\% respectively. Conversely, Boston and Chicago exhibit more substantial gains, reaching up to 11.06\% and 11.96\% respectively. This trend persists with a capacity of 1, revealing higher gains in Boston and Chicago at 14.22\% and 15.35\%, compared to Atlanta, Houston, and LA at 7.55\%, 3.98\%, and 5.56\% respectively. The results underscore that the multi-modal-only setting modestly outperforms both the ride-sharing-only approach and the integrated setting, especially at smaller fleet sizes. However, the service rate rapidly saturates due to the infeasibility of the multi-modal option for the remaining requests. Notably, Boston (64.02\%) and Chicago (52.03\%) yield higher service rates under the multi-modal setting, aligning with their more substantial increase in service rate within the integrated system.

By enabling multi-modal traveling, the system is able to serve 43\% of the demand in both Boston and Chicago with a small fleet of ride-sharing vehicles consisting of only 10 vehicles per 1000 requests. Additionally, it also draws 43\% of the commuters that are currently unable to use transit directly to mass transit increasing the ridership.

Column B of Figure \ref{fig:results} illustrates the composition of service options within the integrated system. With low ride-sharing fleet sizes, the multi-modal option predominates in servicing most requests across all cities. This prioritization stems from the system's emphasis on requests that minimally impact ride-sharing fleet mileage, and multi-modal trips utilize mass transit to cover portions of the journey. However, as the fleet size increases, the proportion of on-demand-only trips gradually supersedes multi-modal trips. This shift is more pronounced for the capacity 1 fleet compared to the capacity 4 fleet, owing to the former's delayed overall service rate.

Column C of Figure \ref{fig:results} analyzes the distribution of multi-modal trips according to their reliance on the ride-sharing fleet for the first and last miles of the journey. These are categorized as 1) both-miles trips, encompassing ride-sharing for both origin-to-departure and arrival-to-destination segments, 2) first-mile-only trips, where ride-sharing is employed for the initial leg, and 3) last-mile-only trips, representing the converse of first-leg trips. Notably, at lower fleet sizes, a significant portion of multi-modal trips (up to 83\%) fall into the last-mile-only category. However, as the fleet size expands, the prevalence of first-mile and both-miles trips increases, while the percentage of last-leg-only trips diminishes. Ratios stabilize as the fleet size reaches 20 vehicles per 1000 requests. Generally, the first mile of a multi-modal journey is more constrained by time compared to the last mile due to the latter's allowance for greater waiting time. Consequently, with lower fleet densities, accommodating both-miles and first-mile trips proves challenging. Similar to the service option composition (Column B), the stabilization of capacity 1 fleets lags behind capacity 4 fleets.

Figure \ref{fig:vmt} offers insight into the total vehicle miles traveled (VMT) if the proposed integrated system was deployed. It is assumed that unserved passengers resort to private vehicles, with their mileage included in the total. The constant line represents total mileage in a scenario where everyone drives. Notably, a capacity 1 ride-sharing fleet exhibits the highest total mileage due to deadheading, making it the least efficient. In Atlanta, Boston, and Chicago, the multi-modal-only approach outperforms both the integrated and ride-sharing-only setups with smaller vehicle fleets, but this advantage diminishes as the fleet size grows. Consistently, the integrated system achieves higher service rates with reduced VMT compared to the ride-sharing-only setup. In specific cases, the integrated approach reduces VMT by up to 20\% in Boston and Chicago, relative to the ride-sharing-only approach. This VMT reduction is less pronounced in Atlanta, Houston, and LA, registering up to 7.7\%, 3.5\%, and 5.5\% respectively. This variation mirrors the earlier observed differences in service rate improvement. As expected, capacity 4 fleets outperform capacity 1 fleet. Implementation of the integrated system offers potential VMT savings of up to 48.7\% (152,222 km) in Chicago with a fleet size of 40 vehicles/1000 requests. Notably, given the 10\% data sampling, these saved VMT figures could escalate to around 1.52 million kilometers (932,000 miles) with 100\% of requests. Most importantly, a significant VMT reduction can be achieved with relatively modest vehicle fleets. In Chicago, a fleet size of 316 vehicles (5 vehicles/1000 requests) in a multi-modal-only setup can reduce 47,554 km (15.2\% of VMT). Similarly, in Boston, a fleet of 224 vehicles can curtail 29,527 km (14.6\% of VMT) of commuter-driven mileage. 

It becomes apparent that notable disparities exist in terms of service rate enhancements and vehicle miles traveled (VMT) reductions among cities when adopting either a multi-modal-only or an integrated transportation system. The city of Chicago demonstrates the most favorable outcomes in both performance metrics, closely trailed by Boston. Conversely, Atlanta, Houston, and Los Angeles demonstrate comparable outcomes amongst themselves, albeit falling significantly behind the favorable benchmarks set by Chicago and Boston.

Figure \ref{fig:transit_only} describes the percentage of commuter trips that can be exclusively (via transit-only option) accommodated by mass transit, taking into consideration the maximum allowable walking distance to transit stops. Notably, the observed percentage distribution across cities closely mirrors the previously identified pattern (Boston and Chicago significantly outperform other cities), indicating that better mass transit leads to an effective multi-modal transportation system.

\section{Conclusions}
We introduce a framework for operating a transit integrated ride-sharing service. Our approach generalizes a state-of-the-art ride-sharing framework to generate feasible travel options, calculate routes, and assign riders to trips. We showcase the benefits of an integrated system through simulations covering five major cities. At urban-scale, we are able to simulate a fleet of up to 2500 ride-sharing vehicles, finding efficient multi-modal routes and the corresponding vehicle assignment for each batch of requests in real-time. We show that the transit-integrated ride-sharing system can decrease the total vehicle miles traveled by the ride-sharing vehicle fleet by up to 20\% while increasing the service rate by up to 12\%. Finally, the framework can be used as an evaluation tool for designing new fixed transit lines for integrated or conventional transit systems and transit-related policy.

\section{Acknowledgment}


This research is partially sponsored by the National Science Foundation under the grant CNS-1952011.









\bibliographystyle{elsarticle-harv} 
\bibliography{mybibfile}

\end{document}